\documentclass[prd,aps]{revtex4}
\usepackage{amsmath}

\begin{document}

\title{Small Brane Black Holes in Randall-Sundrum type I Scenario}
\author{D. Karasik}
\email{karasik@physics.uc.edu}
\author{C. Sahabandu}
\author{P. Suranyi}
\author{L. C. R. Wijewardhana}
\affiliation{Department  of Physics,
University of Cincinnati, Cincinnati, Ohio}

\begin{abstract}
An approximation method to study the properties of a small black
hole located on the TeV brane in the Randall-Sundrum type I scenario is
presented. The method enables us to find the form of the metric
close to the matter distribution when its asymptotic form is given.
The short range solution is found as an expansion in the
ratio between the Schwarzschild radius
of the black hole and the
curvature length of the bulk.
Long range properties are introduced using
the linearized gravity solution as an asymptotic
boundary condition.
The solution is found up to first order. It is valid in the
region close to the horizon but is not valid on the horizon.
The regularity of the horizon is still under study.
\end{abstract}
\pacs{04.50.+h, 04.70.Bw, 04.70.Dy}
\maketitle

\section{Introduction}
Within the Randall-Sundrum brane world
scenarios\cite{rs1,rs2} the four dimensional universe is
viewed as a three-brane with brane tension $\sigma$, embedded in a
AdS five-dimensional bulk, with a cosmological constant $\Lambda$.
A flat brane is achieved by fine-tuning the brane tension and
the cosmological constant
\begin{equation}
     8\pi G_{5}\Lambda=-\frac{6}{\ell^{2}}\;\; ;\;\;
     8\pi G_{5}\sigma=\pm\frac{6}{\ell}~,
     \label{finetune}
\end{equation}
where $\ell$ is the curvature length of the AdS bulk.

   The Randall-Sundrum I (RSI) scenario
was postulated \cite{rs1} to solve the hierarchy problem by
making gravity strong at the weak scale (of a TeV).
  In this scenario the world  consists of two opposite tension branes
embedded in an ADS bulk with   an orbifold structure. The negative
tension brane is called the TeV brane, while the positive tension
brane is called the Planck brane. $Z_{2}$ orbifold symmetry is
assumed about each brane. The ratio between the TeV electro-weak
scale and the Planck scale is taken to be $\lambda\sim10^{-16}$. In
this scenario the standard model particles are assumed to
propagate only on the TeV brane. Gravity, on the other hand, is
five dimensional and propagates in the bulk. The five dimensional
gravitational coupling is of order $1$TeV, $G_{5}=(M_{5})^{-3}$,
$M_{5}\sim 1{\mbox TeV}$. The hierarchy appears on the TeV brane at
distances much larger than the curvature length $\ell$. The
gravitational attraction becomes ordinary four-dimensional
attraction with Newton's constant $G_{4}=(M_{\mbox{Pl}})^{-2}$,
$M_{\mbox{Pl}}=10^{16}{\mbox TeV}$.

   RSII consists of a single positive tension brane  with $Z_2$
symmetry about the brane. This  model does not yield a low scale
gravitational theory.
 From now on  we restrict our attention to the RSI scenario.

Black holes in theories with extra dimensions have been studied
widely.    Myers and Perry~\cite{myers} found Schwarzschild type
solutions (MPS) in  $D$-dimensional asymptotically flat space.  Black hole
solutions were also found in asymptotically AdS space~\cite{hawking1,birmingham}.
No non-trivial closed form  black hole solutions, other than the
black string solution \cite{hawking2}, which extends in a uniform
manner from the brane into the extra dimension,  have been found
in three-brane theories of the Randall-Sundrum type. Given that there is
considerable interest surrounding the production of black holes
at accelerators~\cite{giddings}, and in collisions of cosmic
rays \cite{shapere1},
    it is important to develop approximate methods to find
black hole solutions in Randall-Sundrum brane world theories.

   Some initial attempts at finding black hole solutions centered on
deriving the induced metric on the brane by solving the
Hamiltonian constraint conditions~\cite{induced}. Some of the
induced solutions do not arise from  matter distributions confined
to the brane~\cite{kanti}. Linearized solutions about RS
backgrounds~\cite{lin1} as well as numerical  solutions~\cite{num}
have also been derived. In a recent paper, Casadio and Mazzacurati
~\cite{casadio} investigated how solutions of the Einstein
equation  in RSII models propagate into the bulk by using a
asymptotic expansion of metric coefficients in
   $r^{-1}$,
where $r$ is the radial distance on the brane. They reduced the
Einstein equations to a set of differential equations involving
the expansion coefficients which are functions of the bulk
coordinate $w$. Solving these equations they found a family of
solutions parameterized by two physical parameters, the mass $M$ and
the post-Newtonian parameter $\eta$.

It would be interesting and important to find black hole
solutions in RSI models with TeV scale gravity. We are especially
interested in how such solutions  evolve when the black hole mass
increases starting from around a TeV (where the flat space   MPS
limit is valid) to macroscopic scales.   From the point of view of
cosmology it is useful to
  understand how TeV scale primordial black holes accrete or radiate
matter  when immersed in a  hot plasma. For that we need to work
out their thermodynamic properties. That requires a detailed
determination of the area of  the event horizon as a function of
mass.
  As the size of the black hole approaches the AdS radius, which is of
the same order as the length of separation of the two
branes~\cite{shapere1}, the character of the solution should change
and Gregory-Laflamme type instabilities~\cite{gregory1} could
materialize. One has to find bulk profiles of TeV mass black holes
to address these issues.

A TeV black hole solution in RSI scenario is characterized by two
scales: The curvature length of the bulk $\ell$, generated by the
bulk cosmological constant and the five dimensional Schwarzschild
radius, $\mu$, related to the mass of the black hole as
$\mu=\sqrt{8G_{5}M/(3\pi)}$. For a reasonable setup
\cite{shapere2} the curvature length in RSI is of order
$\ell\sim10({\mbox{TeV}})^{-1}$, while the "mass" can range from a
TeV black hole $\mu\sim1({\mbox{TeV}})^{-1}$, to ordinary black
holes $\mu\sim 10^{16}({\mbox{TeV}})^{-1}$. Clearly, one should
use an appropriate approximation method for various values of the
parameters $\ell, \mu$.

Linearized gravity is valid when the scales under study are much
larger than the Schwarzschild radius $r\gg\mu$. Einstein equations
are expanded to first order in $\mu^{2}\sim G_{5}M$, and the
solution appears to be a superposition of various modes; the zero
mode, the massive modes, and the radion. The radion appears to be
a four dimensional massless mode which describes the relative
motion of the branes. Being massless, the radion mediates a long
range strong gravitational force and therefore ruins the
hierarchy. Provided a stabilization mechanism which in general
makes the radion massive, the linearized solution demonstrates the
transition between five dimensional TeV scaled gravity, at
distances smaller than the curvature length, and four dimensional
Planck scale gravity at distances larger than the curvature length
along the TeV brane. The advantage of the asymptotically
conformally flat linearized solution is to enable the definition
of conserved momenta and identification of mass (in analog to the
ADM mass). On the other hand, one is unable to study the horizon
of the black hole using linearized gravity, since the horizon is
located approximately at $r\sim\mu$. Linearized gravity in RSI
brane scenario was widely studied \cite{rs1,rs2,sv,D,rubakov} and
will be presented in appendix \ref{linearized}.
Asymptotic expansion in $r^{-1}$ \cite{casadio} is valid only when $r\ll\ell,\mu$.
In cases where $\mu<\ell$ the expansion can not be used for $r<\ell$,
and will not enable us to study the horizon.

In this paper we use the following method:
we expand the solution in the dimensionless
parameter $\epsilon=\mu/\ell\ll1$. The coordinates are
rescaled by $\mu$, therefore, an expansion in $\epsilon$ actually
means an expansion in inverse powers of $\ell$. Zeroth order in
$\epsilon$ should be nothing but the five dimensional flat
background MPS black hole \cite{myers}. First order in $\epsilon$
"turns on" the brane tension, while second order in $\epsilon$ "turns
on" the bulk cosmological constant.

Up to first order in $\epsilon$ ($\epsilon$ solution) the metric
is found in section \ref{expansion}. The $\epsilon$ solution is
required to be regular everywhere. Especially it should not
include black strings off the brane. Naturally, the
$\epsilon$ solution is valid in the region $r\ll\ell$. It includes
the region close to the horizon which is essential for horizon
studies. However, it is not valid in the asymptotic region
$r\rightarrow\infty$ and one is unable to identify the mass or
satisfy the junction conditions on the second brane. In order to
overcome these subtleties we use the linearized solution as an
asymptotic boundary condition. Linearized gravity is valid in the
region $r\gg\mu$. We are interested in the case $\mu\ll\ell$,
therefore, there is an intermediate region $\mu\ll r\ll\ell$ where
the $\epsilon$ solution and the linearized solution coincide.
In this intermediate range we take the large distance
behavior of the $\epsilon$ solution and compare it to the short
distance behavior of the linearized solution. As a result, the
long range characteristics (mass identification and second brane
conditions) are introduced into our short ranged
$\epsilon$ solution.

In section~\ref{sec:horizon} we analyze the properties of the
horizon. Clearly, the MPS solution, (zero order in $\epsilon$),
includes a regular horizon with a well defined temperature.
The horizon is modified in the
$\epsilon$ solution. Although the
$\epsilon$ solution is valid in the region close to the horizon,
it is not obvious that the new horizon is regular. The $\epsilon$
solution acquires a null surface defined by $g_{tt}=0$. The normal
vector is null and the static Killing vector is null, i.e. the
surface is a Killing Horizon. The surface gravity of the
horizon is not a constant which indicates that the horizon is
singular \cite{wald}. This leaves us with the following
possibilities; (1) The horizon is singular. Which means that there
are no small black holes in Randall-Sundrum scenario. (2) The
$\epsilon$ expansion does not converge on the horizon. It is valid
only at a distance $\epsilon$ away from the horizon, and therefore
one cannot use it to calculate the surface gravity. (3) The
boundary conditions should be changed thus changing the entire
solution. The issue of the regularity of the horizon is under
study and the detailed discussion is left for a future publication.

\section{An Expansion Close to the Horizon}
\label{expansion}
The goal of this paper is to find a method to
explore the horizon of a small black hole in RSI scenario. The
configuration is characterized by two length scales; the five
dimensional Schwarzschild radius $\mu\,(\sim1\mbox{TeV}^{-1})$,
and the AdS curvature length $\ell\,(\sim10\mbox{TeV}^{-1})$
\cite{shapere2}. The only dimensionless parameter is the ratio
between the two $\epsilon=\mu/\ell$. For a small black hole
$\epsilon$ is smaller than $1$ and the metric can be expanded in
$\epsilon$.

The zero order metric is just the MPS solution
\begin{equation}
    d\bar{s}^{2}=
    -\left(1-\frac{\mu^{2}}{\bar{\rho}^{2}}\right)d\bar{t}^{2}
    +\left(\frac{\bar{\rho}^{2}}{\bar{\rho}^{2}-\mu^{2}}
    \right)d\bar{\rho}^{2}
    +\bar{\rho}^{2}d\psi^{2}
    +\bar{\rho}^{2}\sin^{2}\psi d\Omega_{2}^{2}~.
    \label{ansatz0}
\end{equation}
Where we use spherical coordinates in the $(4+1)$-dimensional bulk
$(\bar{t},\bar{\rho},\psi,\theta,\phi)$. For convenience we
rescale the radial coordinate $\bar{\rho}$ and the time coordinate
$\bar{t}$ by $\mu$ thus making them dimensionless
\begin{equation}
    \rho=\frac{\bar{\rho}}{\mu}\;;\;t=\frac{\bar{t}}{\mu}\;;\;
    g_{AB}=\mu^{2}\bar{g}_{AB}
        \label{scaling}
\end{equation}
The TeV brane is located at $\psi=\pi/2$ and the direction perpendicular
to the brane is $\psi=0$. The Planck brane is located at
$\rho\cos\psi\sim1/\epsilon$. It is outside of the region where the $\epsilon$
expansion is valid.
In general, after rescaling the coordinates by
$\mu$, the ansatz for the metric is
\begin{eqnarray}
    ds^{2}&=&\frac{1}{(1-\epsilon\rho\cos\psi)^{2}}\left[
    -\left(1-\frac{1}{\rho^{2}}-\sum_{n=1}B_{n}(\rho,\psi)\epsilon^{n}\right)dt^{2}
    +\left(\frac{\rho^{2}}{\rho^{2}-1}
    +\sum_{n=1}A_{n}(\rho,\psi)\epsilon^{n}\right)d\rho^{2}
    \right.\nonumber\\
    & & \left.
    +2\sum_{n=1}V_{n}(\rho,\psi)\epsilon^{n} d\rho d\psi
    +\rho^{2}\left(1+\sum_{n=1}U_{n}(\rho,\psi)\epsilon^{n}\right)d\psi^{2}
    +\rho^{2}\sin^{2}\psi d\Omega_{2}^{2}\right]~.
    \label{ansatz}
\end{eqnarray}
The conformal factor in front of the metric is just the
Randall-Sundrum conformal factor and was introduced for
convenience. The rescaled equations of motion include the Einstein
equations \cite{notation}
\begin{equation}
    R_{AB}-\frac{1}{2}R\,g_{AB}=6\epsilon^{2}g_{AB}
    \label{resEinstein}
\end{equation}
and the rescaled Israel junction conditions \cite{junction} on the
TeV brane
\begin{equation}
    2\left(K\gamma_{\mu\nu}-K_{\mu\nu}\right)=6\epsilon\gamma_{\mu\nu}
    \label{resIsrael}
\end{equation}
The Planck brane is located at a distance $\sim\ell$ from the point mass
and therefore the junction condition on the Planck brane can not be evaluated.

Clearly, zeroth order in $\epsilon$ corresponds to vacuum
Einstein equations with no branes. The zeroth order solution is
therefore the MPS black hole \cite{myers}. First order in
$\epsilon$ turns the branes {\em on} but the cosmological constant
of the bulk is still absent. This case corresponds to an empty bulk
with a brane and a point like mass on the brane, it is the main
subject of this paper. Second order in $\epsilon$ turns on the
bulk cosmological constant.

\subsection{First Order in $\epsilon$}
The bulk equations (\ref{resEinstein}) in first order in $\epsilon$
are still the vacuum Einstein equations. However, the
solution will be different from the MPS solution since the brane
tension is of order $\epsilon$ and it will effect the solution
through the junction condition (\ref{resIsrael}).

The ansatz for the metric, up to first order in $\epsilon$, is
\begin{eqnarray}
    ds^{2}=\frac{1}{(1-\epsilon\rho\cos\psi)^{2}}& &\left[
    -\left(1-\frac{1}{\rho^{2}}-\epsilon B_{1}\right)dt^{2}
    +\left(\frac{\rho^{2}}{\rho^{2}-1}
    +\epsilon A_{1}\right)d\rho^{2}\right.\nonumber\\& &
    \left.\frac{}{}+2\epsilon V_{1} d\rho d\psi
    +\rho^{2}\left(1+\epsilon U_{1}\right)d\psi^{2}
    +\rho^{2}\sin^{2}\psi d\Omega_{2}^{2}\right]~.
    \label{ansatz1}
\end{eqnarray}
We choose to work in a gauge where $t$, $\theta$, $\phi$ are kept
unchanged, and $g_{\theta\theta}=\rho^{2}\sin^{2}\psi$. A
convenient gauge in the bulk equations is
\begin{equation}
    V_{1}=\frac{2\rho^{2}-1}{4\rho}A_{1,\psi}
    +\frac{\rho^{2}\cot\psi}{2}U_{1,\rho}~.
    \label{Vgauge}
\end{equation}
The gauge (\ref{Vgauge}) can be achieved using a coordinates
transformation of order $\epsilon$
\begin{equation}
    \rho\rightarrow\rho(1-\epsilon F)\;;\qquad
    \psi\rightarrow\psi+\epsilon F \tan\psi~.
    \label{gaugeF}
\end{equation}
The bulk equations are indeed invariant under the transformation
(\ref{gaugeF}). However, the junction conditions are not invariant
and the function $F(\rho,\psi)$ is not
completely arbitrary. It is subject to junction conditions on the brane
\begin{equation}
    F(\rho,\frac{\pi}{2})=0\;\;;\;\; F_{,\psi}(\rho,\frac{\pi}{2})=0~.
    \label{Fjunction}
\end{equation}
Using the gauge (\ref{Vgauge}), the bulk equations can be solved
in terms of two functions. The gauge function $F(\rho,\psi)$ and
the wave function $H(\rho,\psi)$. The solution is
\begin{subequations}
\label{finalmetric}
\begin{eqnarray}
    B_{1}&=&-\frac{2}{\rho^{3}}\left[
    6\rho H_{0}(\rho)-(2\rho^{2}-1)H_{0}'(\rho)+\rho F(\rho,\psi)\right]~,
    \label{gtt}\\
    A_{1}&=&8\rho^{3}\cos\psi
    -\frac{4\rho^{2}\left(3H(\rho,\psi)+\tan\psi H_{,\psi}(\rho,\psi)\right)}
    {\rho^{2}-1}
    +\frac{12\rho^{2}(\rho^{2}-2)H_{0}(\rho)}{(\rho^{2}-1)^{2}}
    +\frac{2\rho(2\rho^{4}-1)H_{0}'(\rho)}{(\rho^{2}-1)^{2}}
    \nonumber\\& &
    -\frac{2\rho^{2}(2\rho^{2}-1)H_{0}''(\rho)}{\rho^{2}-1}
    +\frac{2\rho^{2}(\rho^{2}-2)F(\rho,\psi)}{(\rho^{2}-1)^{2}}
    +\frac{2\rho^{3}F_{,\rho}(\rho,\psi)}{\rho^{2}-1}~,
    \label{grhorho}\\
    U_{1}&=&
    \frac{2\tan^{2}\psi}{\rho}\left\{6\rho\left[H(\rho,\psi)-H_{0}(\rho)
    \frac{}{}\right]
    -(2\rho^{2}-1)\left[H_{,\rho}(\rho,\psi)-H_{0}'(\rho)\frac{}{}\right]
    \frac{}{}
    -\rho F(\rho,\psi)-\rho\cot\psi F_{,\psi}(\rho,\psi)\right\}~,
    \label{gpsipsi}\\
    V_{1}&=&
    -2\rho^{2}(2\rho^{2}-1)\sin\psi-(4\rho^{2}-1)\tan\psi H_{0}'(\rho)
    +\rho(2\rho^{2}-1)\tan\psi H_{0}''(\rho)\frac{}{}
    +2\rho^{2}\tan\psi H_{,\rho}(\rho,\psi)
    \nonumber\\& &
    +\frac{\rho(2\rho^{2}-1)\tan^{2}\psi H_{,\psi}(\rho,\psi)}{\rho^{2}-1}
    -\rho^{2}\tan\psi F_{,\rho}(\rho,\psi)+\frac{\rho^{3}F_{,\psi}(\rho,\psi)}
    {\rho^{2}-1}~,
    \label{grhopsi}
\end{eqnarray}
\end{subequations}
where $H_{0}(\rho)=H(\rho,\frac{\pi}{2})$.

The wave function $H(\rho,\psi)$ is subject to the differential equation
\begin{equation}
    (\rho^{2}-1)\left(H_{,\rho\rho}-\frac{1}{\rho}H_{,\rho}\right)+H_{,\psi\psi}
    +2\frac{\cos^{2}\psi+1}{\sin\psi\cos\psi}H_{,\psi}=0~.
    \label{Hequation}
\end{equation}
The general solution can be casted as a combination of Associated Legendre functions
in $\rho$ and Hypergeometric functions in $\psi$
\begin{eqnarray}
    H(\rho,\psi)&=&\int\,d\lambda R(\rho;\lambda)\Psi(\psi;\lambda)\label{Hlambda}\\
    R(\rho;\lambda)&=&\rho\sqrt{\rho^{2}-1}\left[a(\lambda)
        Q^{1}_{(\lambda-1)/2}(2\rho^{2}-1)+b(\lambda)
        P^{1}_{(\lambda-1)/2}(2\rho^{2}-1)\right]\label{Rlegendre}\\
    \Psi(\psi;\lambda)&=&c(\lambda)\left._{2}F_{1}(\frac{1-\lambda}{2},\frac{1+\lambda}{2},
        \frac{5}{2},\sin^{2}\psi)\right.+\frac{d(\lambda)}{\sin^{3}\psi}\left.
        _{2}F_{1}(\frac{-2-\lambda}{2},\frac{-2+\lambda}{2},
        -\frac{1}{2},\sin^{2}\psi)\right.\label{Psilambda}
\end{eqnarray}
The first hypergeometric function in Eq.(\ref{Psilambda}) can be
expanded by elementary functions as
\begin{subequations}
\label{Flambda02}
\begin{eqnarray}
    _{2}F_{1}(\frac{1-\lambda}{2},\frac{1+\lambda}{2},\frac{5}{2},\sin^{2}\psi)
    &=&\frac{3[\lambda\cos(\lambda\psi)\sin(2\psi)-
    2\cos(2\psi)\sin(\lambda\psi)]}{2\lambda(4-\lambda^{2})\sin^{3}\psi}
    \label{Flambda}\\
    _{2}F_{1}(\frac{1}{2},\frac{1}{2},\frac{5}{2},\sin^{2}\psi)
    &=&\frac{3[\sin(2\psi)-2\psi\cos(2\psi)]}{8\sin^{3}\psi}
    \label{Flambda0}\\
    _{2}F_{1}(-\frac{1}{2},\frac{3}{2},\frac{5}{2},\sin^{2}\psi)
    &=&\frac{3[4\psi-\sin(4\psi)]}{32\sin^{3}\psi}
    \label{Flambda2}
\end{eqnarray}
\end{subequations}
The function $H(\rho,\psi)$ is subject to the following boundary conditions.

{\it No black strings.} The only source is the
(point) mass on the brane at $\psi=\pi/2$. Therefore, the solution
must be regular everywhere. In particular one should eliminate the
black string which lies in the bulk on a line perpendicular to the
brane at $\psi=0$. The place to look for a string like singularity
is the Kretchmann scalar ($R_{ABCD}R^{ABCD}$). The Kretchmann
scalar depends only on the combination $3H+\tan\psi H_{,\psi}$.
Using the expansion (\ref{Psilambda}) one finds that
\begin{equation}
    3\Psi(\psi;\lambda)+\tan\psi\Psi_{,\psi}(\psi;\lambda)
    =\frac{d(\lambda)(\lambda^{2}-4)}{\psi}+{\cal O}(\psi^{0})~.
\end{equation}
A solution free of black strings requires
\begin{equation}
    d(\lambda)=d_{2}\delta(\lambda-2)\label{d2}~.
\end{equation}
The hypergeometric function associated with $d_{2}$ is simply
\begin{equation}
    \left._{2}F_{1}\left(\frac{-2-\lambda}{2},\frac{-2+\lambda}{2},
        -\frac{1}{2},\sin^{2}\psi\right)\right|_{\lambda=2}=1~.
\end{equation}

{\it Compatibility with Linearized Gravity.} At large distances,
$\rho\rightarrow\infty$, the solution is given by Linearized
Gravity which is an expansion in the mass, $M\sim\mu^{2}$, to
first order. Iterating the linearized solution to include higher
orders of $M$ one finds that only integer powers of $M$ (even
powers of $\mu$) enter the metric. Recalling that
$\epsilon=\mu/\ell\sim M^{1/2}$ and $\rho=\bar{\rho}/\mu\sim
M^{-1/2}$, it is clear that $\epsilon\rho^{-n}\sim M^{(n+1)/2}$.
Since we allow only for integer powers of $M$, the functions
$H(\rho,\psi)$ and $F(\rho,\psi)$ must be odd functions of $\rho$.
The function $R(\rho;\lambda)$ [Eq.(\ref{Rlegendre})] has the same
parity as $\rho^{1\pm\lambda}$. Therefore, the integral over
$\lambda$ must turn into a sum over even values of $\lambda$.

{\it Junction conditions on the brane.} To first order in
$\epsilon$, Israel junction conditions on the TeV brane are simply
\begin{equation}
    B_{1,\psi}(\rho,\frac{\pi}{2})=0;\;\;
    A_{1,\psi}(\rho,\frac{\pi}{2})=0;\;\;
    V_{1}(\rho,\frac{\pi}{2})=0~.
\end{equation}
Using the function (\ref{Hlambda}), these conditions imply that
\begin{equation}
    \int\,d\lambda R(\rho;\lambda)c(\lambda)\cos(\frac{\pi\lambda}{2})
    =-\frac{2}{3}(\rho^{3}-\rho)\label{Junc0}
\end{equation}
The asymptotic behavior of the Legendre functions of the first kind is:
$\rho\sqrt{\rho^{2}-1}P^{1}_{(\lambda-1)/2}(2\rho^{2}-1)\sim \rho^{1+\lambda}$.
Therefore, to prevent higher powers of $\rho$, one must eliminate the Legendre
functions of the first kind except for $\lambda=2$ which is the only function
that can generate an asymptotic behavior $\rho^{3}$.
In particular
\begin{equation}
    b(\lambda)=b_{2}\delta(\lambda-2)\label{b2}
\end{equation}
Still, the Legendre function of the first kind, with $\lambda=2$,
includes a term $(\log\rho)/\rho$ in the asymptotic expansion. In
order to eliminate the logarithmic term from the condition
(\ref{Junc0}) one must use the combination
\begin{equation}
    \left[P^{1}_{(\lambda-1)/2}(2\rho^{2}-1)
    +\frac{4}{\pi^{2}}\frac{\partial}{\partial\lambda}Q^{1}_{(\lambda-1)/2}
    (2\rho^{2}-1)\right]_{\lambda=2}~.
\end{equation}
This combination can be obtained by setting
\begin{equation}
    a(\lambda)=\sum_{n=0}a_{2n}\delta(\lambda-2n)
    -\frac{4}{\pi^{2}}b_{2}\delta'(\lambda-2)\label{alambda}.
\end{equation}

{\it Asymptotic boundary conditions.} One would like to impose
boundary conditions at $\rho\rightarrow\infty$. The problem is
that the $\epsilon$ expansion is valid only for
$1\leq\rho\ll1/\epsilon$ (which corresponds to
$\mu\leq\bar{\rho}\ll\ell$). However, the asymptotic form of the
metric is given by the linearized solution. The later is valid for
$1\ll\rho$ ($\mu\ll\bar{\rho}$). As long as $\epsilon\ll1$, there
is an intermediate region $1\ll\rho\ll1/\epsilon$ where both the
$\epsilon$ solution and the linearized solution are valid.
Therefore, the large $\rho$ limit of the $\epsilon$ solution
should coincide with first order in $\epsilon$ of the linearized
solution. The appropriate asymptotic boundary condition is
\begin{equation}
    \lim_{\rho\rightarrow\infty}\rho(g^{\epsilon}_{AB}-g^{L}_{AB})=0~.
    \label{asympcond}
\end{equation}

To summarize, collecting Eqs.(\ref{Hlambda}), (\ref{d2}), (\ref{b2}), and (\ref{alambda}),
the function $H(\rho,\psi)$ is
\begin{eqnarray}
    H(\rho,\psi)&=&\rho\sqrt{\rho^{2}-1}\left[a_{0}Q^{1}_{-1/2}(2\rho^{2}-1)
    \frac{3[\sin(2\psi)-2\psi\cos(2\psi)]}{8\sin^{3}\psi}\right.\nonumber\\
    & &\left.+Q^{1}_{1/2}(2\rho^{2}-1)\left(a_{2}\frac{3[4\psi-\sin(4\psi)]}
    {32\sin^{3}\psi}
    +\frac{d_{2}}{\sin^{3}\psi}
    -b_{2}\frac{3[8\psi+4\psi\cos(4\psi)-3\sin(4\psi)]}{32\pi^{2}\sin^{3}\psi}
    \right)\right.\nonumber\\
    & &\left.
    +b_{2}\left(P^{1}_{1/2}(2\rho^{2}-1)+\left.\frac{2}{\pi^{2}}
    \frac{\partial Q^{1}_{n-1/2}(2\rho^{2}-1)}{\partial n}\right|_{n=1}\right)
    \frac{3[4\psi-\sin(4\psi)]}{32\sin^{3}\psi}\right.\nonumber\\
    & &\left.
    +\sum_{n=2}a_{2n}Q^{1}_{n-1/2}(2\rho^{2}-1)
    \frac{3[n\cos(2n\psi)\sin(2\psi)-
    \cos(2\psi)\sin(2n\psi)]}{8n(1-n^{2})\sin^{3}\psi}\right]~.
    \label{Hdiscrete}
\end{eqnarray}
It includes a discrete set
of parameters; $d_{2}$, $b_{2}$, and $a_{2n}\;(n=0,1,..)$.
These parameters should be fixed by the asymptotic boundary condition
(\ref{asympcond}) and the junction condition (\ref{Junc0}).

So, we turn first to the asymptotic behavior of the solution.
The linearized solution appears in appendix~\ref{linearized}.
Expansion of the linearized solution to first
order in $\epsilon$ appears in Eqs.(\ref{gL}). The latter imply
\begin{subequations}
\label{gLepsilon}
\begin{eqnarray}
    B_{1}^{L}&=&\frac{1}{\rho}\left[\cos\psi+\frac{\psi}{\sin\psi}\right]~,
    \label{B1L}\\
    A_{1}^{L}&=&\frac{1}{\rho}\left[-\frac{5}{4}\cos\psi
    -\frac{3}{2}\psi\sin\psi+\frac{\psi}{4\sin\psi}
    -\rho\cos\psi W_{1,\rho}\right]~,
    \label{A1L}\\
    U_{1}^{L}&=&
    \frac{1}{\rho}\left[\frac{11}{8}\cos\psi+\frac{5}{8}\cos(3\psi)
    -\frac{2\psi}{\sin\psi}+\frac{3}{2}\psi\sin\psi
    +\sin\psi W_{1,\psi}\right]~,
    \label{U1L}\\
    V_{1}^{L}&=&
    \frac{\cos^{2}\psi(2-5\cos(2\psi))}{8\sin\psi}
    +\frac{3\psi\cos(3\psi)}{8\sin^{2}\psi}
    +\frac{1}{2}\rho\sin\psi W_{1,\rho}
    -\frac{1}{2}\cos\psi W_{1,\psi}~.
    \label{V1L}
\end{eqnarray}
\end{subequations}
The function $W_{1}(\rho,\psi)$ is
arbitrary, but is subject to the boundary conditions
\begin{equation}
    W_{1}(\rho,\frac{\pi}{2})=0\;\;;\;\;W_{1,\psi}(\rho,\frac{\pi}{2})=\frac{\pi}{4}
    ~.\label{W1Lconditions}
\end{equation}
To simplify the analysis [to avoid the function
$W_{1}(\rho,\psi)$], one should look at the following
combinations; $B_{1}$,
$C_{1}=\left(A_{1}/(\cos\psi)\right)_{,\psi}
    +\left(\rho U_{1}/(\sin\psi)\right)_{,\rho}$,
    $D_{1}=\sin^{2}\psi A_{1}+\cos^{2}\psi U_{1}
    +V_{1}\sin(2\psi)/\rho$, and
    $\left.U_{1}^{L}\right|_{\psi=\pi/2}$.
\begin{subequations}
\label{gLcomb}
\begin{eqnarray}
    &B_{1}^{L}&=-\frac{1}{\rho}\left[\cos\psi+\frac{\psi}{\sin\psi}\right]~,
    \label{B1Lcomb}\\
    &C_{1}^{L}
    &=\frac{1}{\rho\sin^{2}(2\psi)}\left[2\psi\cos(2\psi)-\sin(2\psi)
    +\frac{3}{4}\sin(4\psi)-3\psi\right]~,
    \label{Lcons1}\\
    &D_{1}^{L}&=
    \frac{5}{16\rho\sin\psi}\left[\sin(4\psi)-4\psi\right]~,
    \label{Lcons2}\\
    &\left.U_{1}^{L}\right|_{\psi=\pi/2}&=0~.
    \label{Lcons3}
\end{eqnarray}
\end{subequations}
For the purpose of comparing the $\epsilon$ solution and the
linearized solution, the $\epsilon$ solution should be expanded in
inverse powers of $\rho$ only up to order $\epsilon\rho^{-1}\sim
M$. Expanding the solution (\ref{finalmetric}, \ref{Hdiscrete}) to
first order in $\rho^{-1}$, and casting the gauge function as
$F(\rho,\psi)=\rho F_{1}(\psi)+F_{2}(\psi)/\rho$, the asymptotic
behavior of the $\epsilon$ solution is
\begin{subequations}
\label{gepsilonasymp}
\begin{eqnarray}
    &B_{1}^{\epsilon}&=\frac{1}{\rho}\left[\frac{3\pi^{2}}{4}a_{0}-2F_{1}(\psi)\right]~,
    \label{B1asymp}\\
    &C_{1}^{\epsilon}&=
    \frac{1}{\rho\sin^{2}(2\psi)}\left[\frac{}{}8\sin^{3}\psi F_{1}(\psi)
    +8\cos\psi\sin^{2}\psi F_{1}'(\psi)\right.\nonumber\\& &\left.-
    3\pi a_{0}\left(3\psi\cos(2\psi)-3\sin\psi\cos\psi+\pi\sin^{3}\psi\right)
    +\frac{9 b_{2}}{4\pi}\left(\sin(4\psi)-4\psi\right)\right]~,
    \label{econs1}\\
    &D_{1}^{\epsilon}&=
    \frac{4}{\pi}\rho\cos\psi\sin^{2}\psi\left[\pi-3b_{2}\right]
    +\frac{1}{\rho}\left[\frac{3\pi a_{0}}{4\sin\psi}\left(
    \psi-2\psi\cos(2\psi)+\sin\psi\cos\psi-\pi\sin^{3}\psi\right)\right.\nonumber\\&
    &\left.
    +\frac{3b_{2}}{8\pi}\left(\cos\psi+5\cos(3\psi)-\frac{6\psi}{\sin\psi}\right)
        +2\sin^{2}\psi F_{1}(\psi)+\sin(2\psi)F_{1}'(\psi)\right]~,
    \label{econs2}\\
    &\left.U_{1}^{\epsilon}\right|_{\psi=\pi/2}&=
    \rho\left[\frac{3\pi^{2}a_{0}}{8}+F_{1}''(\pi/2)\right]+
    \frac{1}{\rho}\left[-\frac{3\pi^{2}a_{0}}{32}-
    \frac{27\pi^{2}a_{2}}{64}-\frac{9\pi d_{2}}{4}
    -\frac{45b_{2}}{16}+F_{2}''(\pi/2)\right]~.
    \label{econs3}
\end{eqnarray}
\end{subequations}
Comparing $B_{1}^{L}$ and $B_{1}^{\epsilon}$ [Eqs.(\ref{B1Lcomb}), (\ref{B1asymp})]
and using the junction condition for $F(\rho,\psi)$ [Eq.(\ref{Fjunction})]
one finds that
\begin{equation}
    a_{0}=-\frac{2}{3\pi}\;\;;\;
    F_{1}(\psi)=\frac{1}{2}\left(\cos\psi+\frac{\psi}{\sin\psi}
        -\frac{\pi}{2}\right)~.
    \label{a0F1}
\end{equation}
Going next to $D_{1}^{L}$ and $D_{1}^{\epsilon}$ [Eqs. (\ref{Lcons2}), (\ref{econs2})].
One finds that
\begin{equation}
    b_{2}=\frac{\pi}{3}~.
    \label{b2value}
\end{equation}
Finally, Eqs.(\ref{Lcons3}) and (\ref{econs3}) imply that
\begin{equation}
    F_{2}''(\frac{\pi}{2})=\frac{\pi}{64}(56+144d_{2}+27\pi a_{2})~.
    \label{F2''}
\end{equation}
The rest of the set $a_{2n}$ can be evaluated by means of
the junction condition on the brane (\ref{Junc0})
\begin{eqnarray}
    -\frac{2}{3}(\rho^{3}-\rho)&=&
    \sum_{n=1}a_{2n}(-1)^{n}\rho\sqrt{\rho^{2}-1}Q^{1}_{n-1/2}(2\rho^{2}-1)
        -\frac{2}{3\pi}\rho\sqrt{\rho^{2}-1}Q^{1}_{-1/2}(2\rho^{2}-1)\nonumber\\& &
    -\frac{\pi}{3}\rho\sqrt{\rho^{2}-1}\left(P^{1}_{1/2}(2\rho^{2}-1)
    +\left.\frac{2}{\pi^{2}}
    \frac{\partial Q^{1}_{n-1/2}(2\rho^{2}-1)}{\partial n}\right|_{n=1}\right)~.
    \label{Junc1}
\end{eqnarray}
To simplify Eq.(\ref{Junc1}), we set $z=2\rho^{2}-1$ and divide
the equation by $\rho\sqrt{\rho^{2}-1}$
\begin{equation}
    \sum_{n=1}a_{2n}(-1)^{n}Q^{1}_{n-1/2}(z)=
    -\frac{2}{3}\sqrt{\frac{z-1}{2}}
    +\frac{2}{3\pi}Q^{1}_{-1/2}(z)
    +\frac{\pi}{3}P^{1}_{1/2}(z)
    +\frac{2}{3\pi}
    \frac{\partial}{\partial n} Q^{1}_{n-1/2}(z)|_{n=1}~.
    \label{Junc0z}
\end{equation}
according to a theorem by
Banerjee~\cite{banerjee} the toroidal functions $Q^{1}_{n-1/2}(z)$
form a complete set on $(1,\infty)$.
The function on the right hand side goes to zero as
$\rho\rightarrow\infty$. Therefore it can be expanded in a series of toroidal
functions of the second kind i.e. the left hand side. The
prescription given by Banerjee~\cite{banerjee} is quite complicated
and we will not follow it here. Alternatively, the coefficients can be found
recursively from the asymptotic expansion of Eq.(\ref{Junc1})
\begin{equation}
    a_{2}=-\frac{1}{3\pi}\;;\;a_{4}=\frac{4}{9\pi}\;;\;a_{6}=\frac{41}{21\pi}
    \;\ldots
\end{equation}
A single parameter $d_{2}$ remains undetermined.

\section{The Black Hole Horizon}
\label{sec:horizon}
The existence of an event horizon is a delicate issue. It is clear
that before turning on the brane ($\epsilon=0$), there exists a
single horizon at $\rho_{H}=1$. There is no guarantee that first
order in $\epsilon$ will not ruin the horizon.

We assume that the new horizon is located at $\rho_{H}=1+\epsilon\zeta$.
We expand
$g_{tt}(\rho=1+\epsilon\zeta)=0$ to first order in $\epsilon$ to get
\begin{equation}
    2\epsilon\zeta+2\epsilon\left[F(1,\psi)+1
    +\beta\log|\frac{\epsilon\zeta}{8}|\right]=0~,
    \label{horizoneq}
\end{equation}
with
$\beta=(6d_{2}+1)/8$
depend on the parameter $d_{2}$.

The gauge function $F(1,\psi)$ introduces a $\psi$ dependence.
But, there is still a gauge freedom to choose $F(1,\psi)=0$ which
is in complete agreement with the junction conditions
(\ref{Fjunction}) and (\ref{a0F1},\ref{F2''}). We choose the gauge
$F(1,\psi)=0$ from now on.

The logarithmic term comes from the associated Legendre functions of the
second kind which close to $1$ behave like
\begin{equation}
    \rho\sqrt{\rho^{2}-1}Q^{1}_{n-1/2}(2\rho^{2}-1)\sim
    -\frac{1}{2}+\frac{(4n^{2}-1)(\rho-1)}{4}\left[-\log|\rho-1|+C_{n}\right]
    +{\cal O}((\rho-1)^{2})~.
\end{equation}
This is the first indication that the solution (\ref{Hdiscrete})
is not regular on the horizon. The second indication comes from the surface gravity.
The surface gravity calculated from the solution (\ref{Hdiscrete}) is not constant
on the horizon. According to \cite{wald} this shows that the horizon is singular.

There are few possible explanations. One option is that the
configuration indeed becomes singular already at first order in
$\epsilon$. This means that there are no small black holes in
Randall-Sundrum scenario. Another possibility is that although the
$\epsilon$ solution is valid close to the horizon it does not
converge on the horizon. This can be seen if one tries to expand
$g_{tt}^{-1}$ in $\epsilon$, clearly, the expansion does not
converge on the horizon ($g_{tt}=0$).

The regularity of the horizon is still under study. The detailed discussion
of this issue is left for a future publication.

\section{Summary}
Randall-Sundrum $1$ model makes the gravitational interaction strong
on small scales and might allow for small black holes on the TeV
brane. Interesting phenomena might arise in the presence of TeV
scale black holes. Especially, the way such primordial black-holes
accrete or radiate matter when immersed in a hot plasma would
affect the cosmological evolution of the universe. The behavior of
a growing or evaporating black hole is related to its
thermodynamic properties. Thus the study of its horizon is
required.

The purpose of this paper is to introduce an approximation method
for studying the horizon of a small black hole in the Randall-Sundrum
I scenario. 'Small' black hole means that the five dimensional
Schwarzschild radius of the black hole is smaller than the
curvature length of the bulk AdS manifold. The method suggested in
this paper is to expand the metric in the dimensionless parameter
$\epsilon=\sqrt{G_{5}M}/\ell$, and to fix the asymptotic behavior
using the weak field approximation (linearized gravity).
Zeroth order in $\epsilon$ is just
a Myers-Perry flat background black hole. First order in
$\epsilon$ corresponds to the presence of a brane with brane
tension which is of order $\epsilon$ in a flat bulk. Second order
in $\epsilon$ includes also the bulk cosmological constant which
is of order $\epsilon^{2}$.

We fix the asymptotic behavior in the following way.
The $\epsilon$ solution is valid in the region
$r\ll\ell$. It includes the region close to the horizon which is
essential for horizon studies. However, it is not valid in the
asymptotic region $r\rightarrow\infty$ and one is unable to
identify the mass or satisfy the junction conditions on the second
brane (at distance $\sim\ell$). In order to overcome these
subtleties we use the linearized solution as an asymptotic
boundary condition. Linearized gravity is valid in the region
$r\gg\mu$. We are interested in the case $\mu\ll\ell$, therefore,
there is an intermediate region $\mu\ll r\ll\ell$ where the
$\epsilon$ solution and the linearized solution can be matched. In
this intermediate range we take the large distance behavior of the
$\epsilon$ solution and match it to the short distance behavior of
the linearized solution. As a result, the long range
characteristics (mass identification and second brane conditions)
are introduced into the $\epsilon$ solution.

We should mention that the method of expanding in $\epsilon$ and
comparing with the linearized solution as an asymptotic condition is
applicable to other scenarios as well. For example: Randall-Sundrum
single brane scenario, de Sitter bulk
(expanding in the ratio between the Schwarzschild
radius and the curvature length),
compact extra dimension (expanding in the ratio between the Schwarzschild
radius and the compactification radius), etc.

In this paper we have found a solution up to first order in
$\epsilon$. We require that the curvature singularity resides only
on the brane. No black strings are associated with our solution.
The mass of the object consists of matter completely confined
to the brane.

Although the $\epsilon$ solution is valid in the region close to
the horizon, it is not obvious that the new horizon is regular.
The $\epsilon$ solution acquires a null surface defined by
$g_{tt}=0$. The normal vector is null and the static Killing
vector is null i.e. the surface is a Killing Horizon. But, the
surface gravity of the horizon is not a constant which indicates
that the horizon is singular \cite{wald}. This leaves us with the
following possibilities; (1) The horizon is singular. Which means
that there are no small black holes in Randall-Sundrum scenario.
(2) The $\epsilon$ expansion does not converge on the horizon. It
is valid only at a distance $\epsilon$ away from the horizon, and
therefore one cannot use it to calculate the surface gravity. (3)
The boundary conditions should be changed thus changing the entire
solution.

The issue of the regularity
of the horizon is under study and the detailed discussion is left
for a future publication.

\begin{acknowledgments}
This work is supported in part by the U.S. Department of Energy Grant
No. DE-FG02-84ER40153.
We thank Philip Argyres and Cenalo Vaz for fruitful discussions.
\end{acknowledgments}

\appendix
\section{Linearized Gravity}
\label{linearized}
\subsection{Randall-Sundrum I Background}
RSI scenario consists of two flat 3-branes embedded in five
dimensional AdS bulk. The equations of motion in the bulk are the
five dimensional Einstein equations \cite{notation}
\begin{equation}
    R_{AB}-\frac{1}{2}R\,g_{AB}=8\pi G_{5} T_{AB}=\frac{6}{\ell^{2}}g_{AB}~.
    \label{Einstein}
\end{equation}
With an additional $Z_2$ symmetry with respect to the brane,
the Israel junction conditions are
\begin{equation}
    2\left(K\gamma_{\mu\nu}-K_{\mu\nu}\right)=
    8\pi G_{5}S_{\mu\nu}=\frac{\pm6}{\ell}\gamma_{\mu\nu}~.
    \label{Israel}
\end{equation}
The upper (lower) sign corresponds to the negative (positive) tension brane.

Parameterizing the bulk with the coordinates $y^{A}=(x^{\mu},w)$,
where $w$ is the coordinate perpendicular to the brane, a
conformally flat ansatz for the metric is
\begin{equation}
    ds_{5}^{2}=a^{2}(w)\eta_{AB}dy^{A}dy^{B}~.
    \label{RSmetric}
\end{equation}
We choose a coordinate system in which
the TeV brane (with negative tension) is located
at $w=0$, while the Planck brane (positive tension) is at $|w|=\ell(1-\lambda)$.
$\lambda$ is a very small number. It is the ratio between the TeV
scale and the Planck scale, namely $\lambda\sim10^{-19}$.
The conformal factor in (\ref{RSmetric}) is
$a^{2}(w)=\left(\frac{\ell}{\ell-|w|}\right)^{2}$.
It has the value $1$ on the TeV brane (at $w=0$)
and the value $\lambda^{-2}$ on the Planck brane.
In this scenario the standard model particles are assumed to be
confined by some mechanism to the TeV brane.

Let us assume now that there is a matter distribution on the
brane such that it is confined to a small region on the brane at
$w=0$, therefore, the energy momentum tensor is non vanishing only
at $(r<r_{0},w=0)$. However, if we are interested in the solution
far from this region, we can use the linearized
version of Eqs.(\ref{Einstein},\ref{Israel}). The perturbed metric
is given by
\begin{equation}
    ds_{5}^{2}=a^{2}(w)
    \left[\eta_{AB}+h_{AB}(x^{\mu},w)\right]dy^{A}dy^{B}~.
    \label{bulkmetric}
\end{equation}
The perturbation $h_{AB}$ is assumed to be small. The source for
the perturbation is the matter on the brane, characterized by the
energy momentum tensor, $T_{\mu\nu}$. The total energy momentum
tensor on the brane (which enters Israel junction condition (\ref{Israel}))
is $S_{\mu\nu}=\sigma\gamma_{\mu\nu}+T_{\mu\nu}$.

\subsection{Conserved Momentum}
The definition of a conserved momentum is usually done in an
asymptotically Minkowski $n+1$ dimensional space-time as
$P^{A}=\int d^{n}x\, T^{A0}$. It is associated with energy
momentum conservation $ T^{AB}_{\;,B}=0$, where the use of
ordinary derivative instead of covariant derivative is essential.

In RS models, one deals with asymptotically {\em conformal}
Minkowski space-time, and the energy-momentum conservation
condition is $ \left.T^{AB}\right._{;B}=0$. One can verify,
however, that the energy momentum conservation condition can be
written as
\begin{equation}\label{EMcon}
     T_{A\;;B}^{\,B}=\frac{1}{\sqrt{-g}}(\sqrt{-g}T_{A}^{\,B})_{,B}
     -\frac{a'(w)}{a(w)}T_{B}^{B}\delta_{Aw}=0~.
\end{equation}
Therefore, since the conformal factor $a(w)$ depends only on the
fifth coordinate, one is able to define a conserved four momentum
$P^{(5)}_{\mu}=\int d^{3}x\,dw\,\sqrt{-g} T_{\mu}^{0}$. One can
verify that the momentum is conserved using Eq.(\ref{EMcon})
\begin{equation}
    \dot{P}^{(5)}_{\mu}=-\int d^{3}x\,dw\,[(\sqrt{-g}T_{\mu\;}^{i})_{,i}
    +(\sqrt{-g} T_{\mu\;}^{w})_{,w}]=-\lim_{r\rightarrow\infty}\int dw\,d\Omega\, r^{2}n_{i}\sqrt{-g}
    T_{\mu\;}^{i}-\int d^{3}x\,dw\,(\sqrt{-g}T_{\mu\;}^{w})_{,w}~.
\label{Momcon}
\end{equation}
The first integral on the right hand side
vanishes provided the matter is confined to a
small region. The second integral on the right hand side
is treated differently in
different scenarios; in RSII, with a single brane, the $w$
integration is over the interval $(-\infty,\infty)$ and vanishes
provided the matter is confined to the brane and the energy-momentum component
$ T_{\mu}^{w}$ is continuous. In RSI with two branes the only
requirement is the continuity of $T_{\mu}^{w}$.

It is worth noting that if the energy-momentum tensor is
traceless (for example four dimensional Maxwell field),
then one is able to define a conserved five momentum $P^{(5)}_{A}=\int
d^{3}x\,dw\,\sqrt{-g} T_{A}^{0}$. In other words, the presence of
the trace brakes a symmetry and therefore the theory has only
a conserved four momentum.

The solution to Eq.(\ref{EMcon}) is given by $\sqrt{-g}
T_{\mu}^{B}=\partial_{D}Q^{DB}_{\;\mu}$, with
$Q^{DB}_{\;\mu}=-Q^{BD}_{\;\mu}$. The exact form of $Q^{DB}_{\;\mu}$
can be calculated from the linearized Einstein equations.
The conserved momentum can be calculated using
\begin{equation}\label{Momentum}
    P^{(5)}_{\mu}=\int d^{3}x\,dw\,\partial_{w}Q^{w0}_{\mu}+\int
    d\Omega\,dw\, r^{2}n_{i}Q^{i0}_{\mu}~.
\end{equation}
The first integral has contributions only from the discontinuities of
$Q^{w0}_{\mu}$. This can be evaluated using
the linearized junction conditions and turns out to be exactly the four
momentum of matter confined to the branes
\begin{equation}\label{Momentum54}
\int d^{3}x\,dw\,\partial_{w}Q^{w0}_{\mu}=\int
d^{3}x\,\left.T_{\mu}^{0}\right|_{\mbox{branes}}=P^{(4)}_{\mu}~.
\end{equation}
The five dimensional momentum can be summarized as
\begin{eqnarray}
    P^{(5)}_{\mu}=P^{(4)}_{\mu}&+&\lim_{r\rightarrow\infty}\frac{1}{16\pi G_{5}}\int
    d\Omega\,dw\, a^{3}(w)r^{2}n^{i}(-h_{\mu i,0}-h^{0}_{\mu,i}
    +\eta_{\mu i}h^{0\nu}_{,\nu}-\delta^{0}_{\mu}h^{j}_{i,j}
    +\delta^{0}_{\mu}h^{B}_{B,i}+\eta_{\mu i}h^{B}_{B,0})~.
\label{Momentum5}
\end{eqnarray}
The energy $P_{0}$ is given by
\begin{equation}\label{Energy5}
    P^{(5)}_{0}=P^{(4)}_{0}+\lim_{r\rightarrow\infty}\frac{1}{16\pi G_{5}}\int
    d\Omega\,dw\, a^{3}(w)r^{2}n^{i}[-h_{0 i,0}-h^{j}_{i,j}
    +h^{k}_{k,i}+h_{ww,i}]~.
\end{equation}
If the only sources for the perturbation reside on the branes,
then the five dimensional and the four dimensional momenta are
equal, and the integrals in Eqs.(\ref{Momentum5},\ref{Energy5})
must vanish.
\subsection{Linearized
Solution} Choosing the gauge $g_{\theta\theta}=a^{2}(w)r^{2}$, the
metric perturbation caused by a single point-like mass, $M$,
located on the TeV brane is given in radial coordinates as
\begin{subequations}
\label{hI}
\begin{eqnarray}
    h_{tt}(r,w)&=&\frac{2C}{r}+\frac{2G_{5}M}{3r\ell}\left[2\tilde{F}_{1}(\alpha,\beta)
    \right]  +2\frac{W(r,w)}{\ell-w}~,
    \label{h00I}\\
    h_{rr}(r,w)&=&\frac{2C}{r}+\frac{2G_{5}M}{3r\ell}\left[\tilde{F}_{1}(\alpha,\beta)
    -\alpha \tilde{F}_{1,\alpha}\right]+\frac{2rW_{,r}(r,w)}{\ell-w}~,
    \label{hrrI}\\
    h_{\theta\theta}(r,w)&=&0~,
    \label{httI}\\
    h_{ww}(r,w)&=&-2\left(\frac{W(r,w)}{\ell-w}+W_{,w}(r,w)\right)~,
    \label{hwwI}\\
    h_{rw}(r,w)&=&\frac{G_{5}M}{3\ell^{2}}
    \tilde{F}_{2}-W_{,r}(r,w)+\frac{r}{\ell-w}
    \left(\frac{W(r,w)}{\ell-w}+W_{,w}(r,w)\right)~.
    \label{hrwI}
\end{eqnarray}
\end{subequations}
The term $\frac{2C}{r}$ is the contribution of the zero-mode. The constant $C$ cannot
be determined by the bulk equations nor by the junction conditions. The only way
to determine $C$ would be through the definition of the total mass of the
configuration given by Eq.(\ref{Energy5}).

The terms with $\tilde{F}_{1},\tilde{F}_{2}$ are the massive
tachyonic modes of the linearized equations. These functions
depend on the dimensionless variables $\alpha=r/\ell$,
$\beta=(\ell-w)/\ell$, and are given by
\begin{subequations}
\label{tildeF1F2}
\begin{eqnarray}
       \tilde{F}_{1}(\alpha,\beta)&=
    \frac{2\beta^{2}}{\pi}&\int_{0}^{\infty}dz\,\sin(z\alpha)
    \frac{K_{1}(z\lambda)I_{2}(z\beta)+I_{1}(z\lambda)K_{2}(z\beta)}
    {K_{1}(z\lambda)I_{1}(z)-I_{1}(z\lambda)K_{1}(z)}~,
    \label{tildeF1}\\
    \tilde{F}_{2}(\alpha,\beta)&=
    \frac{2\beta^{2}}{\pi}&\int_{0}^{\infty}dz\,\left[\sin(z\alpha)-\frac{
    \sin(z\alpha)}{z^{2}\alpha^{2}}
    +\frac{\cos(z\alpha)}{z\alpha}\right]z
    \frac{K_{1}(z\lambda)I_{1}(z\beta)-I_{1}(z\lambda)K_{1}(z\beta)}
    {K_{1}(z\lambda)I_{1}(z)-I_{1}(z\lambda)K_{1}(z)}~.
    \label{tildeF2}
\end{eqnarray}
\end{subequations}
$W(r,w)$ is a gauge function which cannot be determined by the bulk equations.
It represents a coordinates transformation of the form
\begin{equation}
    r\rightarrow r\;;\qquad
    w\rightarrow w+W(r,w)
    \label{Wgauge}
\end{equation}
It is subject to the following boundary conditions
\begin{eqnarray}\label{WIconditions}
    W(r,w=0)=\frac{G_{5}M}{3r}\;&;&\;W_{,w}(r,w=0)=-\frac{G_{5}M}{3r\ell}\nonumber\\
    W(r,w=\ell(1-\lambda))=0\;&;&\;W_{,w}(r,w=\ell(1-\lambda))=0
\end{eqnarray}
On the TeV brane ($w=0$) $W(r,0)$ behaves like a massless four dimensional scalar
mode, and is called "the radion".

The definition of the mass and the hierarchy should be seen at the asymptotic region
$r\gg\ell$. The asymptotic expansion of the functions $\tilde{F}_{1},\tilde{F}_{2}$ is
given by
\begin{subequations}
\label{largealpha}
\begin{eqnarray}
    \tilde{F}_{1}(\alpha\gg1,\beta)&\sim& \frac{2\lambda^{2}}{1-\lambda^{2}}~,\\
    \tilde{F}_{2}(\alpha\gg1,\beta)&\sim& -\frac{\beta(\beta^{2}-\lambda^{2})}
    {\alpha^{2}(1-\lambda^{2})}~.
\end{eqnarray}
\end{subequations}
Using Eq.(\ref{Energy5}) and assuming that there are no other sources besides the
point mass on the TeV brane, one can fix the constant $C$ in the zero mode
$C=-\frac{2G_{5}M\lambda^{2}}{3\ell(1-\lambda^{2})}$. The zero
mode tends to be repulsive.

The gauge invariant force acting on a slowly moving test particle
is given by
\begin{equation}
    F_{r}=\frac{1}{2}h_{tt,r}+\frac{1}{\ell-w}h_{rw}
    \label{force}
\end{equation}
The gravitational potential along the TeV brane at distances much larger
than the curvature length $r\gg\ell$ is given by
\begin{equation}
    V=-\frac{2G_{5}M\lambda^{2}}{3r\ell(1-\lambda^{2})}-\frac{G_{5}M}{3r\ell}
    \label{potential4I}
\end{equation}
The first term is the contribution of the zero-mode and the massive modes.
These sum up together to give the correct hierarchy with
$G_{4}=\frac{2G_{5}\lambda^{2}}{3\ell(1-\lambda^{2})}$.

The last term in Eq.(\ref{potential4I}) is the contribution of the radion
and is clearly destroying the hierarchy and producing a long range strong
(TeV) gravitational attraction. In order to save the hierarchy it is necessary
to include a stabilization mechanism that removes the radion
at large distances. A stabilization mechanism was suggested in
\cite{GW}. It introduces a bulk scalar field which is coupled to the
radion by some potential and cause the radion to acquire mass. Thus, the
radion acquires a
gravitational potential of the form $e^{-m_r r}/r$ and does not contribute
at distances $r\gg\frac{1}{m_r}\log(\frac{1}{\lambda^2})$.
In general, the introduction of a stabilizing bulk scalar
field will change the RSI background, namely the conformal factor, and might have
an impact on the graviton modes as well.

In this paper we are interested in the
short range region $\mu<r<\ell$. Therefore, assuming that the radion becomes massive
by some mechanism but is still light compared to the curvature length
$m_r<\frac{1}{\ell}\sim 10-100GeV$,
we can assume that in the region $r<\ell$ the above discussion is valid and the
metric for the linearized solution is given by Eqs.(\ref{hI}). In addition,
we assume that the stabilizing bulk scalar field is massive enough not to interfere
in the region $\mu<r<\ell$, thus assuming $M_{\Phi}>\frac{1}{\mu}\sim1TeV$.

The asymptotic expansion of the functions $\tilde{F}_{1},\tilde{F}_{2}$ in
the region $r\ll\ell$ is given by
\begin{subequations}
\label{smallalpha}
\begin{eqnarray}
    \tilde{F}_{1}(\alpha\ll1,\beta)&\sim& \frac{2\alpha\beta^{3/2}}
    {\pi(\alpha^{2}+(1-\beta)^{2})}-\frac{3(5-\beta)\sqrt{\beta}}{4\pi}
    \arctan(\frac{\alpha}{1-\beta}) ~,   \\
    \tilde{F}_{2}(\alpha\ll1,\beta)&\sim&\frac{4\alpha(1-\beta)\beta^{3/2}}
    {\pi(\alpha^{2}+(1-\beta)^{2})^{2}}
    +\frac{(3\beta^{2}+2\beta+3)\sqrt{\beta}}{4\pi\alpha}\left[
    \frac{1-\beta}{\alpha^{2}+(1-\beta)^{2}}
    -\frac{1}{\alpha}\arctan(\frac{\alpha}{1-\beta})\right]~.
\end{eqnarray}
\end{subequations}
Using these expansions, and also expanding the gauge function as
\begin{equation}
    W(r,w)=\frac{\mu^{2}}{2}\left[W_{0}(r,w)+\frac{1}{\ell}W_{1}(r,w)\right]
    \label{W01}
\end{equation}
one can write the expanded form of the metric (\ref{hI}), to first
order in $1/\ell$.
\begin{subequations}
\label{hell}
\begin{eqnarray}
    h_{tt}(r,w)&=&\frac{\mu^{2}}{r^{2}+w^{2}}+\frac{\mu^{2}}{\ell}\left[
    -\frac{3}{2r}\arctan(\frac{r}{w})-\frac{3w}{2(r^{2}+w^{2})}+W_{0}(r,w)\right]~,
    \label{h00ell}\\
    h_{rr}(r,w)&=&\frac{\mu^{2}r^{2}}{(r^{2}+w^{2})^{2}}+\frac{\mu^{2}}{\ell}\left[
    -\frac{3}{4r}\arctan(\frac{r}{w})+\frac{3w(w^{2}-r^{2})}{4(r^{2}+w^{2})^{2}}
    +r W_{0,r}(r,w)\right] ~,  \label{hrrell}\\
    h_{\theta\theta}(r,w)&=&0~,
    \label{httell}\\
    h_{ww}(r,w)&=&-\mu^{2}W_{0,w}(r,w)-\frac{\mu^{2}}{\ell}\left[\frac{}{}
    W_{0}(r,w)+W_{1,w}(r,w)\right]~,
    \label{hwwell}\\
    h_{rw}(r,w)&=&\mu^{2}\left[-\frac{1}{4r^{2}}\arctan(\frac{r}{w})
    +\frac{w(3r^{2}+w^{2})}{4r(r^{2}+w^{2})^{2}}-\frac{1}{2}W_{0,r}(r,w)\right]\nonumber\\&
    &    +\frac{\mu^{2}}{\ell}\left[\frac{3w}{8r^{2}}\arctan(\frac{r}{w})
    -\frac{3w^{2}(3r^{2}+w^{2})}{8r(r^{2}+w^{2})^{2}}+\frac{r}{2}W_{0,w}(r,w)
    -\frac{1}{2}W_{1,r}(r,w)\right]~.
    \label{hrwell}
\end{eqnarray}
\end{subequations}
The boundary conditions for $W(r,w)$ (\ref{WIconditions}) are
\begin{eqnarray}
    W_{0}(r,0)=\frac{\pi}{4r} &\;\;;\;\;& W_{1}(r,0)=0\;; \nonumber\\
    W_{0,w}(r,0)=0 &\;\;;\;\;& W_{1,w}(r,0)=-\frac{\pi}{4r}~.
    \label{W01conditions}
\end{eqnarray}
In order to compare the linearized solution (\ref{hell}) and the
$\epsilon$ expansion (\ref{finalmetric}), one has to perform a
coordinate transformation to rescale the linearized-metric
(\ref{hell})
\begin{equation}
    \rho=\frac{\sqrt{r^{2}+w^{2}}}{\mu}\;\;;\;\;
    \psi=\arctan(\frac{r}{w})\;\;;\;\;
    g_{AB}\rightarrow \frac{1}{\mu^{2}}g_{AB}\;\;;\;\;\epsilon=\frac{\mu}{\ell}~.
    \label{transeL}
\end{equation}
After transformation (\ref{transeL}) the linearized metric (\ref{hell})
is
\begin{subequations}
\label{gL}
\begin{eqnarray}
    (1-\epsilon\rho\cos\psi)^{2}g_{tt}^{L}&=&-1+
    \frac{1}{\rho^{2}}-\frac{\epsilon}{\rho}\left[\cos\psi+\frac{\psi}{\sin\psi}\right]~,
    \label{gttL}\\
    (1-\epsilon\rho\cos\psi)^{2}g_{\rho\rho}^{L}&=&1+
    \frac{1}{\rho^{2}}+\frac{\epsilon}{\rho}\left[-\frac{5}{4}\cos\psi
    -\frac{3}{2}\psi\sin\psi+\frac{\psi}{4\sin\psi}
    -\rho\cos\psi W_{1,\rho}\right]~,
    \label{grhorhoL}\\
    (1-\epsilon\rho\cos\psi)^{2}g_{\psi\psi}^{L}&=&\rho^{2}+
    \epsilon\rho\left[\frac{11}{8}\cos\psi+\frac{5}{8}\cos(3\psi)
    -\frac{2\psi}{\sin\psi}+\frac{3}{2}\psi\sin\psi
    +\sin\psi W_{1,\psi}\right]~,
    \label{gpsipsiL}\\
    (1-\epsilon\rho\cos\psi)^{2}g_{\rho\psi}^{L}&=&\epsilon
    \left[\frac{\cos^{2}\psi(2-5\cos(2\psi))}{8\sin\psi}
    +\frac{3\psi\cos(3\psi)}{8\sin^{2}\psi}
    +\frac{1}{2}\rho\sin\psi W_{1,\rho}
    -\frac{1}{2}\cos\psi W_{1,\psi}\right]~,
    \label{grhopsiL}\\
    (1-\epsilon\rho\cos\psi)^{2}g_{\theta\theta}^{L}&=&\rho^{2}\sin^{2}\psi~.
    \label{gthetathetaL}
\end{eqnarray}
\end{subequations}
The first term of the gauge function $W(\rho,\psi)$ (\ref{W01}) is
fixed by setting $\epsilon=0$
\begin{equation}
    W_{0}(\rho,\psi)=\frac{1}{2\mu\rho}\left(\cos\psi+\frac{\psi}{\sin\psi}\right)~.
    \label{W0fixed}
\end{equation}
The second part of the function $W(\rho,\psi)$ (\ref{W01}) is yet
undetermined, but is subject to the boundary conditions
\begin{equation}
    W_{1}(\rho,\frac{\pi}{2})=0\;\;;\;\;W_{1,\psi}(\rho,\frac{\pi}{2})=\frac{\pi}{4}
    ~.\label{W1conditions}
\end{equation}


\begin{thebibliography}{10}
\bibitem{notation} The notation: bulk indices are upper case Latin
letters, brane indices are Greek letters, $G_{5}$ is Newton's
constant in five dimensions, $\gamma_{\mu\nu}$ is the induced
metric on the brane, and $K_{\mu\nu}$ is the extrinsic curvature
of the brane.
\end{thebibliography}
\end{document}